\documentstyle[nato,numreferences,epsfig]{crckapb}

\newcommand{\cZ}{{\cal Z}}
\newcommand{\cF}{{\cal F}}

\newcommand{\Z}{{Z \!\!\! Z}}
\newcommand{\beqn}{\begin{eqnarray}}
\newcommand{\eeqn}{\end{eqnarray}}

\newcommand{\dd}{\mbox{d}}

\newcommand{\preprint}{\vspace{-13.3cm}
 \rightline{\small ITEP-LAT/2002-01}
 \rightline{\small KANAZAWA-02-03}
 \vspace{1mm}
 \rightline{\small 17 February, 2002}
 \vspace{11.6cm}}
\newcommand{\slovakia}{\vspace{-6mm}
 \leftline{\footnotesize{* Presented by the first author
 at the NATO Advanced Research Workshop}}
 \vspace{-1mm} \leftline{\footnotesize{
 "Confinement, Topology, and other Non-Perturbative Aspects of
 QCD",}}
 \vspace{-1mm} \leftline{\footnotesize{
 Star\'a Lesn\'a, High Tatra Mountains (Slovakia),
 January 21-27, 2002}} \vspace{3mm}}

\newcommand{\eq}[1]{(\ref{#1})}

\begin{opening}
\title{Lattice monopoles in hot SU(2) gluodynamics \protect\\
as blocked continuum defects}
\author{M. N. CHERNODUB${}^{\lowercase{a,b}}$}
\author{K. ISHIGURO${}^{\lowercase{b}}$}
\author{T. SUZUKI${}^{\lowercase{b}}$}
\institute{${}^{a}$ Institute of Theoretical and Experimental Physics\\
B. Cheremushkinskaja, 25, Moscow, 117259, Russia\\
${}^{b}$ Institute of Theoretical Physics,\\
University of Kanazawa, Kanazawa 920-1192, Japan}

\end{opening}

\runningtitle{Blocked monopoles in hot gluodynamics}

\begin{document}
\begin{abstract}
We propose to consider lattice monopoles in gluodynamics as
continuum monopoles blocked to the lattice. In this approach the
lattice is associated with a measuring device consisting of
finite--sized detectors of monopoles (lattice cells). Thus a
continuum monopole theory defines the dynamics of the lattice
monopoles. We apply this idea to the static monopoles in high
temperature gluodynamics. We show that our suggestion allows to
describe the numerical data both for the density of the lattice
monopoles and for the lattice monopole action in terms of a
continuum Coulomb gas model.
\end{abstract}

\slovakia\preprint

\section{Introduction}

One of the most successful approaches to the confinement phenomena
in QCD is the based on the so--called dual superconductor
mechanism~\cite{DualSuperconductor}. The key role in the mechanism
is played by Abelian monopoles which  are identified with the help
of the Abelian projection method~\cite{AbelianProjections} based
on the partial gauge fixing of non--Abelian gauge symmetry up to a
residual Cartan (Abelian) subgroup. The monopoles appear in the
theory due to compactness of the Cartan subgroup. According to the
numerical results~\cite{MonopoleCondensation} the monopoles are
condensed in the low temperature (confinement) phase. The
condensation of the monopoles leads to formation of the
chromoelectric string which implies confinement of color. The
importance of the Abelian monopoles is stressed by the Abelian
dominance phenomena which was first observed in the lattice $SU(2)$
gluodynamics. In the so-called Maximal Abelian projection the
monopoles make a dominant contribution to the
zero temperature string tension~\cite{AbelianDominance}. At high
temperatures, (deconfinement phase) the monopoles are responsible
for the spatial string tension~\cite{AbelianDominanceT}.

We propose to describe the dynamics of the Abelian monopoles in
the $SU(2)$ gluodynamics considering the lattice as a kind of a
"monopole detector". For the sake of simplicity we are working in
the deconfinement phase dominated by static monopoles while
monopoles running in spatial directions are suppressed. We
investigate the physics of the static monopole currents which is
effectively three dimensional.

\section{Ideology}

We suppose that the lattice with a finite lattice spacing $b$ is
embedded in the continuum space-time. Each lattice cell, $C_s$,
detects the total magnetic charge, $k_s$, of "continuum" monopoles
inside it:
\beqn
k_s = \int\limits_{C_s}\dd^3 x\, \rho(x)\,, \qquad
\rho(x) = \sum_a  q_a \, \delta^{(a)} (x - x^{(a)})\,,
\eeqn
where $\rho$ is the density of the continuum monopoles,
$q_a$ and $x_a$ is the position and the charge (in units of
a fundamental magnetic charge, $g_M$) of $a^{\mathrm{th}}$
continuum monopole. We stress the difference between continuum and
lattice monopoles: the continuum monopoles are fundamental
objects while the lattice monopoles are
associated with non--zero magnetic charges of continuum
monopoles located inside lattice cells, $k_s \neq 0$.

Before going into details we mention that our approach is similar
to the blocking of the monopole degrees of freedom from fine to
coarser lattices~\cite{BlockSpin}, which allows to define perfect
quantum actions for topological defects. Another similarity can be
observed with ideas of Ref.~\cite{BlockingOfFields} where the
blocking of the continuum {\it fields} to the lattice was
proposed. Our approach is based on blocking of the continuum {\it
topological defects} to the lattice, and, as a result, is more
suitable for the investigation of the lattice monopoles. Indeed,
blocking of the fields~\cite{BlockingOfFields} leads to {\it
non--integer} lattice magnetic currents which makes a comparison
of the numerical results with the analytical predictions
difficult.

Below we show that properties of the Abelian lattice monopoles --
found in numerical simulations of hot $SU(2)$ gluodynamics -- can
be described by the continuum blocking proposed above. Suppose the
dynamics of the continuum monopoles in the high temperature
gluodynamics is governed by the standard $3D$ Coulomb gas model:
\beqn
\cZ = \sum\limits_{N=0}^\infty \frac{\zeta^N}{N!}
\Biggl[\prod\limits^N_{a=1} \int \dd^3 x^{(a)} \sum\limits_{q_a = \pm 1}\Biggr]
\exp\Bigl\{ - \frac{g^2_M}{2} \sum\limits_{\stackrel{a,b=1}{a \neq b}}^N
q_a q_b \, D(x^{(a)}-x^{(b)})\Bigr\}\,,
\label{CoulombModel}
\eeqn
where $\zeta$ is the so--called fugacity parameter and $D(x)$  is the
inverse Laplacian in continuum. In this simple
model the continuum monopoles are supposed to be point--like
while the monopole core in $SU(2)$ gluodynamics is of a finite
size~\cite{FiniteRadius}. However our results presented below show that
this simplification works at high temperatures.

Below we choose the {\it {v.e.v.}} of the continuum monopole
density, $\rho$, and the Debye mass, $m_D$, as suitable parameters
of the continuum model (instead of $g_M$ and $\zeta$). In the
leading order of the dilute gas approximation we have~\cite{Polyakov}:
$\rho = 2 \zeta$ and $m_D = g_M \sqrt{2 \zeta}$. It is worth mentioning that
the $3D$ Debye mass corresponds to a {\it magnetic} screening mass
in four dimensions.

We are interested in basic quantities characterizing the lattice
mo\-no\-po\-les: the {\it{v.e.v.}} of the squared
magnetic charge, $\langle k^2_s\rangle$, and the
monopole action, $S_{mon}(k)$. In addition to the parameters
of the continuum model both these quantities should depend
on the lattice monopole size, $b$. Note that we study the quantity
$\langle k^2_s\rangle$ instead of the density, $\langle |k_s|\rangle$,
since the analytical treatment of the density is difficult. However
philosophically both these quantities are equivalent.

In the dilute gas approximation we get~\cite{ToBePublished}:
\beqn
\langle k^2_s\rangle(b) = \rho \, b^3 \cdot P(b \slash
\lambda_D,L)\,,
\nonumber
\eeqn
\beqn
P(\mu,L) = \frac{1}{L^3} \, \sum\limits_{\vec q \in \Z}
\frac{{\vec u}^2}{{\vec u}^2 + \mu^2} \, \prod^3_{i=1} {\Biggl[
\frac{2 \sin [u_i(q) \slash 2]}{u_i(q)} \Biggr]}^2\,,\quad
u_i(q) = \frac{2 \pi q_i}{L}\,,
\nonumber
\eeqn
where $L$ is the lattice size. The scale for the monopole size is given by
the (magnetic) Debye screening length, $\lambda_D = 1 \slash m_D$.
In a leading order we get for small and large lattice cubes:
\beqn
\langle k^2_s\rangle(b) =
\left\{
\begin{array}{ll}
\rho \cdot b^3 + \cdots\,, & b \ll \lambda_D\,; \\
C\cdot \rho \, \lambda_D \cdot b^2 + \cdots\,,
& b \gg \lambda_D\,.
\end{array}
\right.
\label{TheorDensity}
\eeqn
This expression is written in a thermodynamic limit where
$C\approx 2.94$.

Analogously we may derive an effective action for the monopoles.
Substituting the unity,
\beqn
1 = \prod\limits_s
\sum\limits_{k_s \in \Z} \delta\Biggl(k_s - \int_{C_s} \dd^3 x\,
\rho(x)\Biggr)\,,
\eeqn
into the Coulomb gas partition function \eq{CoulombModel} and
integrating over all continuum monopoles we get
in the dilute gas approximation~\cite{ToBePublished}:
\beqn
Z & = & \sum_{k \in \Z} e^{- S_{mon}(k)}\,, \,\,\,
S_{mon}(k) = \frac{1}{2 \rho \, b^3} \sum\limits_{s,s'} k_s \,
\cF_{s,s'}(b \slash \lambda_D)\,k_{s'}\,,
\nonumber\\
\cF_{s,s'}(\mu) & = & \frac{1}{L^3} \sum\limits^L_{\vec m =0}\,
\cF^{-1}\left(\frac{2 \pi \vec m}{L}, \mu\right)
\cdot \exp \Bigl\{2 \pi i (\vec s - \vec s\,', \vec m) \slash L\Bigr\}\,,
\nonumber\\
\cF(\vec u, \mu) & = & \sum_{\vec r \in \Z} \sum_{i=1}^3
\frac{4\, \sin^2 (u_i \slash 2)}{(\vec u + 2 \pi \vec r)^2 + \mu^2}\,
\prod^3_{\stackrel{j=1}{j \neq i}} {\left(
\frac{2 \sin (u_j \slash 2)}{u_j + 2 \pi r_j} \right)}^2\,,
\nonumber
\eeqn
where we omit higher--order corrections. According to this
equation the leading order contribution to the monopole action is
quadratic in monopole fields.

Similarly to the density of the squared monopole charge the
monopole action depends on the ratio $b \slash \lambda_D$:
\beqn
S_{mon}(k) =
\left\{
\begin{array}{lll}
\frac{1}{4\rho} \cdot \frac{1}{b^3} \cdot  \sum\limits_{s} k^2_s + \cdots\,, &
b \ll \lambda_D\,; \\
\frac{1}{4\rho\, \lambda_D} \cdot \frac{1}{b^2}
\cdot \sum\limits_{s,s'} k_s\, D_{s,s'}\, k_{s'} + \cdots\,,&
b \gg \lambda_D\,,
\end{array}
\right.
\label{TheorAction}
\eeqn
where $D_{s,s'}$ is the inverse Laplacian on the lattice. Leading
contribution to the monopole action is given by the massive
(Coulomb) terms for small (large) lattice monopoles.

\section{Analytics vs. Numerics}

To study numerically the monopoles of various lattice sizes we use
the extended monopole construction~\cite{ExtendedMonopoles}. The
size of the extended monopole is $b = n a$, where $n$ is the
lattice blocking size and $a$ is the lattice spacing of the fine
lattice on which the fields of the $SU(3)$ gluodynamics are
defined. We study only the static components, $k_4$, of the $4D$
monopole currents, $k_\mu$. The lattice blocking is performed only
in spatial directions, $n_s=1 \dots 8$.  We simulate $SU(2)$
gluodynamics on the lattices $48^3 \times L_t$ with $L_t = 4 \dots
12$ at temperatures $T \slash T_c = 1.6 \dots 4.8$. The size
of the lattice monopoles is measured in terms of the zero
temperature string tension, $\sigma_{T=0}$.

\begin{figure}[!htb]
\begin{center}
\begin{tabular}{cc}
  \epsfig{file=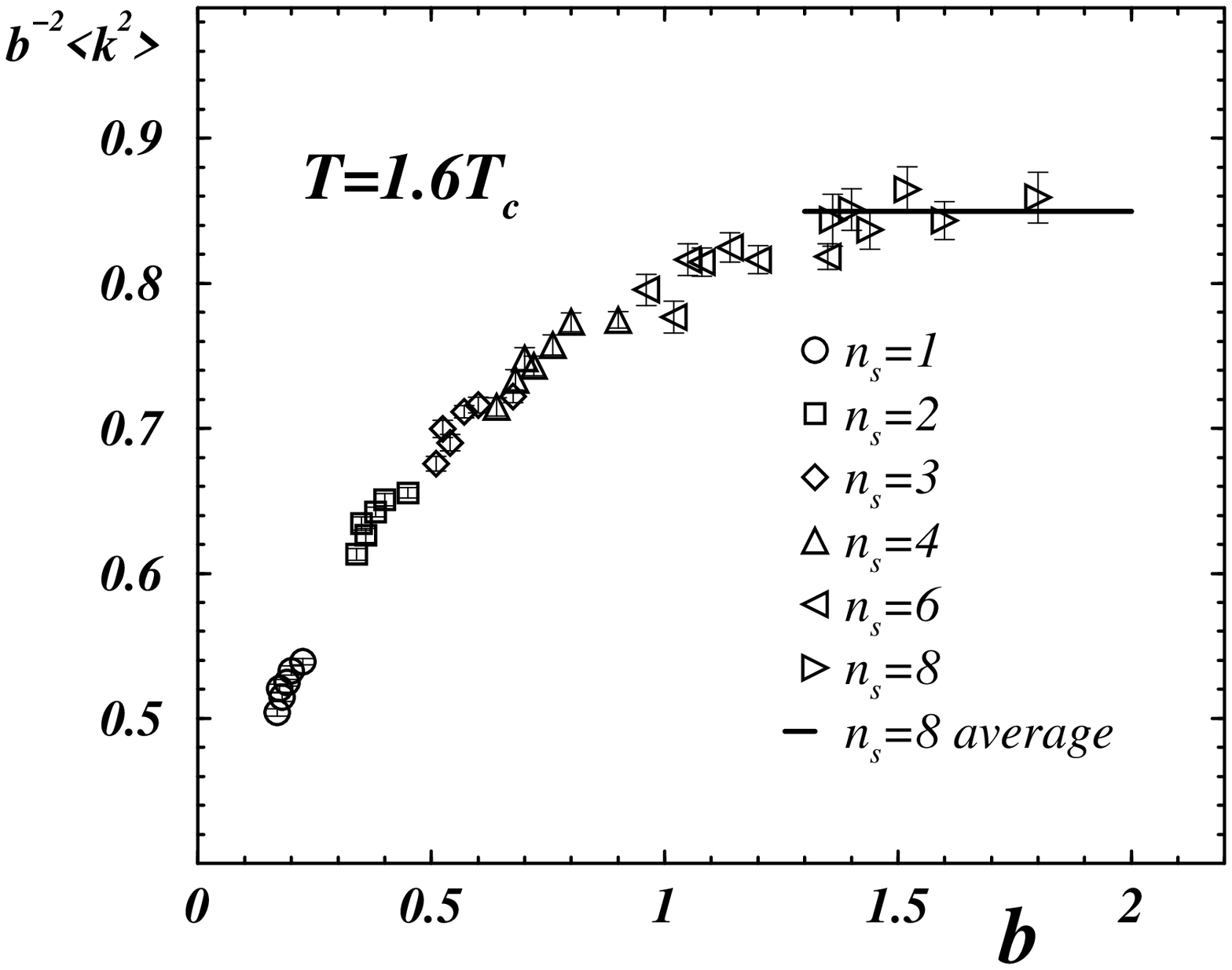,width=5.8cm,height=5.0cm} &
  \epsfig{file=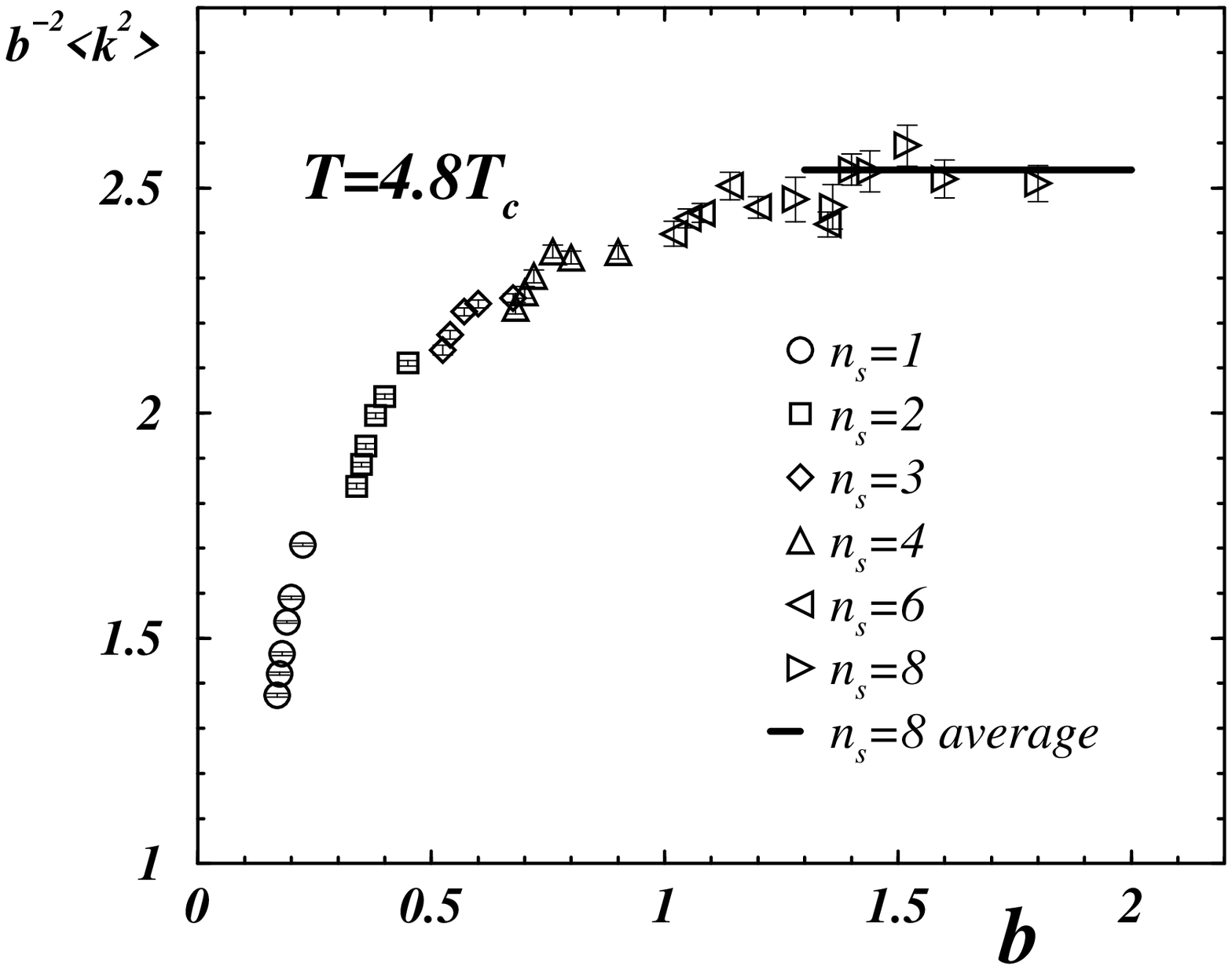,width=5.8cm,height=5.0cm}\\
(a) & (b)\\
\end{tabular}
\end{center}
 \caption{The ratio of the squared magnetic charge density and $b^2$,
 $\langle k^2(b)\rangle \slash b^2$, $vs.$ lattice monopole size, $b$, at
 (a) $T = 1.6 \, T_c$ and  (b) $T = 4.8 \, T_c$.}
\label{fig:p2:mon:dens}
\end{figure}

First we check the density of squared lattice monopole charge
$vs.$ lattice monopole size, $b$, Figure~\ref{fig:p2:mon:dens}.
According to eq.\eq{TheorDensity} the function $\langle
k^2(b)\rangle \slash b^2$ must vanish at small monopole sizes and
tend to constant at large $b$. This behaviour can be observed in
our numerical data, Figure~\ref{fig:p2:mon:dens}, up to some jumps
for densities with different $n_s$. We ascribe these jumps to the
lattice artifacts emerged due to finiteness of the fine lattice
spacing, $a$, and finite volume effects. In
Figure~\ref{fig:p2:mon:dens} the monopole size, $b$, is measured
in units of the zero string tension, $\sigma_{T=0}$.
The large-$b$ asymptotics of $\langle k^2(b)\rangle \slash b^2$ is
approximated by averaging of the appropriate $n_s=8$ data.
According to eq.\eq{TheorDensity} the large-$b$ asymptotics can
be used to extract the dimensionless quantity,
\beqn
R(T) = \frac{\sigma_{T=0}}{\lambda_D(T)\,\rho(T)}\,,
\label{R}
\eeqn
which is discussed in the next Section.

\begin{figure}
\begin{minipage}{12.5cm}
\begin{center}
\begin{tabular}{cc}
  \epsfig{file=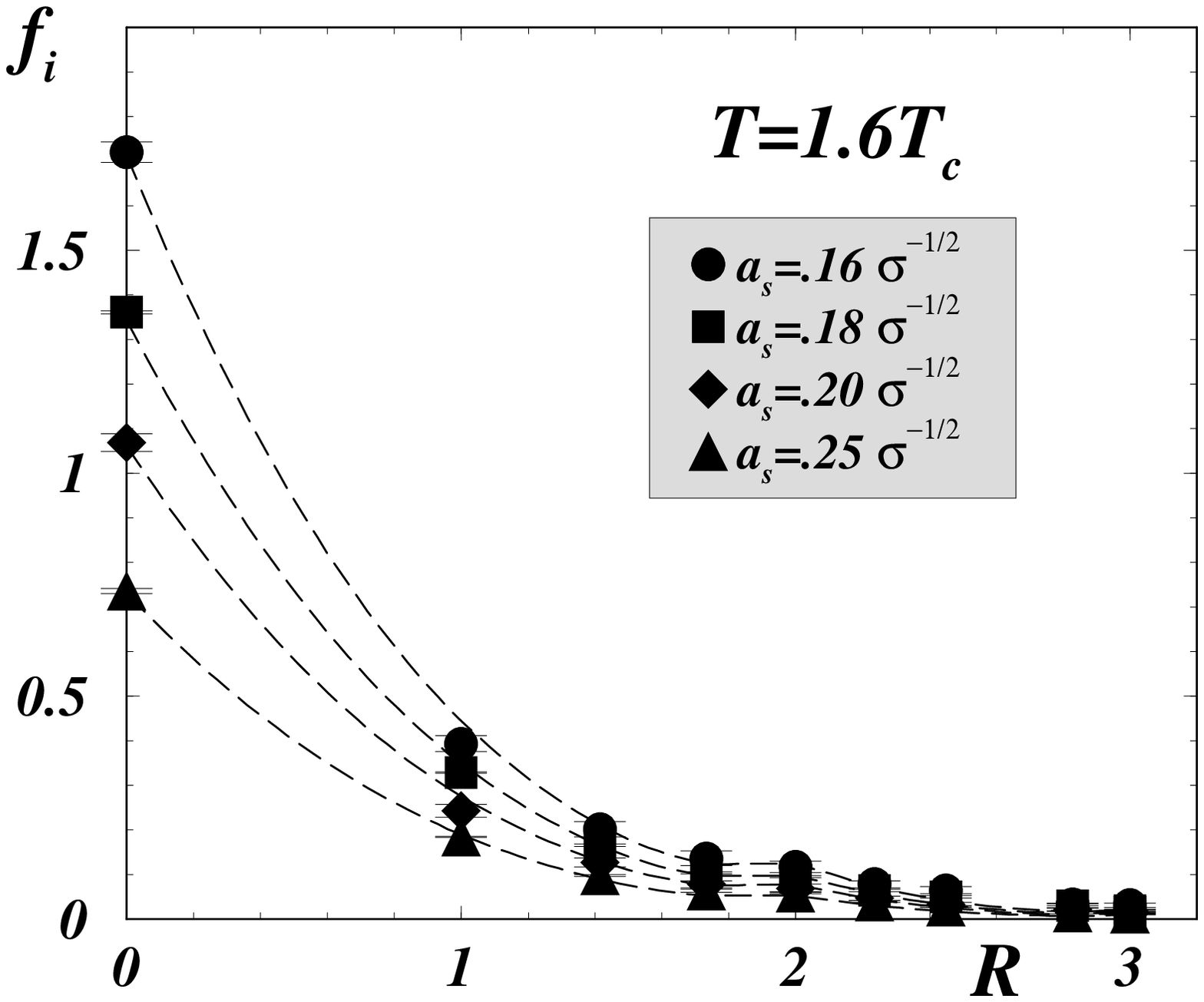,width=5.8cm,height=5.0cm} &
  \epsfig{file=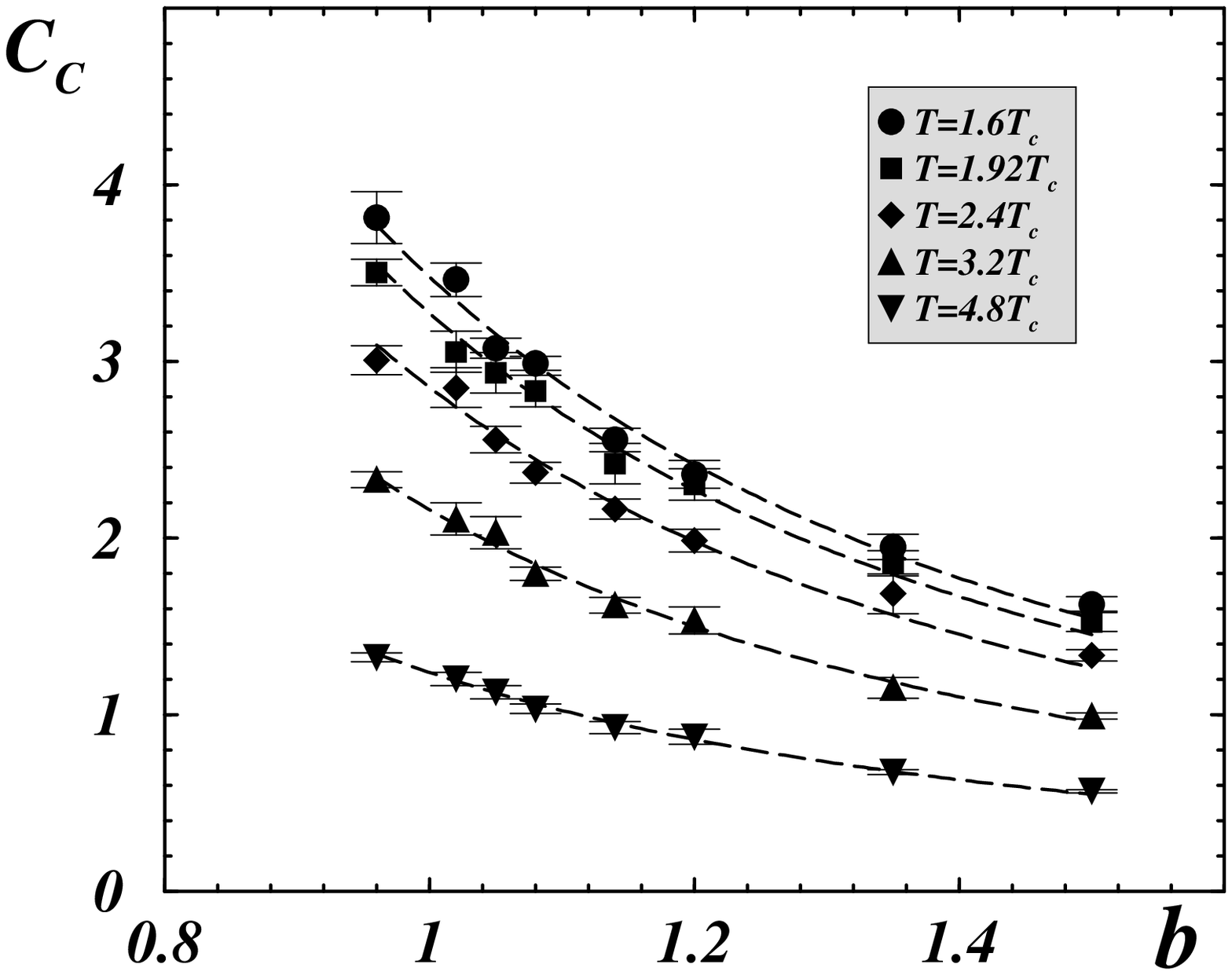,width=5.8cm,height=5.0cm}\\
(a) & (b)\\
\end{tabular}
 \end{center}
 \end{minipage}
\caption{(a) Two--point coupling constants, $f_i$, of the monopole
action $vs.$ the distance between the lattice points, $r$ (in
lattice units). The numerical data corresponds to $T=1.6 T_c$,
$n_s=6$ and various spatial spacings, $a_s$, of the fine lattice.
The fits by the Coulomb interaction~\eq{CoulombLaw} are visualized
by the dashed lines.
(b) The pre-Coulomb coupling $C_C$ and the fits of $C_C$ by
eq.~\eq{PreCoulombAction} are shown for various temperatures, $T$.}
\label{fig:fit:act}
\end{figure}

The monopole action for the static monopole currents, $k_{s}
\equiv k_{s,4}$, at high temperatures was found numerically in
Ref.~\cite{NumericalMonopoleAction}. The leading order terms in
the action are quadratic:
\beqn
S_{mon}(k) = \sum_i f_i \, S_i(k)\,,
\eeqn
where $S_i$ are two--point operators of the monopole charges (see
Ref.~\cite{NumericalMonopoleAction} for details). An example of
the monopole action for certain parameters is shown in
Figure~\ref{fig:fit:act}(a). In the same Figure we show
the fitting of the action by the Coulomb action,
\beqn
S_{mon}(k) = C_C \cdot \sum\limits_{s,s'} k_s\, D_{s,s'}\,
k_{s'}\,,
\label{CoulombLaw}
\eeqn
which follows from eq.\eq{TheorAction}. Here $C_C$ is the fitting
parameter. This {\it one-parametric} fit works almost perfectly.

According to eq.\eq{TheorAction} the pre-Coulomb coefficient
$C_C(b,T)$  at sufficiently large monopole size , $b \gg \lambda_D$,
must scale as follows:
\beqn
C_C(b,T) = R(T) \cdot b^{-2}\,,
\label{PreCoulombAction}
\eeqn
where $R$ is defined in eq.\eq{R}. We present the data for the
pre-Coulomb coefficient and the corresponding {\it one-parameter}
fits~\eq{PreCoulombAction} in Figure~\ref{fig:fit:act}(b). The
agreement between the data and the fits is very good.

\section{Check of Coulomb gas picture}

\begin{figure}[!htb]
\begin{center}
\begin{tabular}{cc}
  \epsfig{file=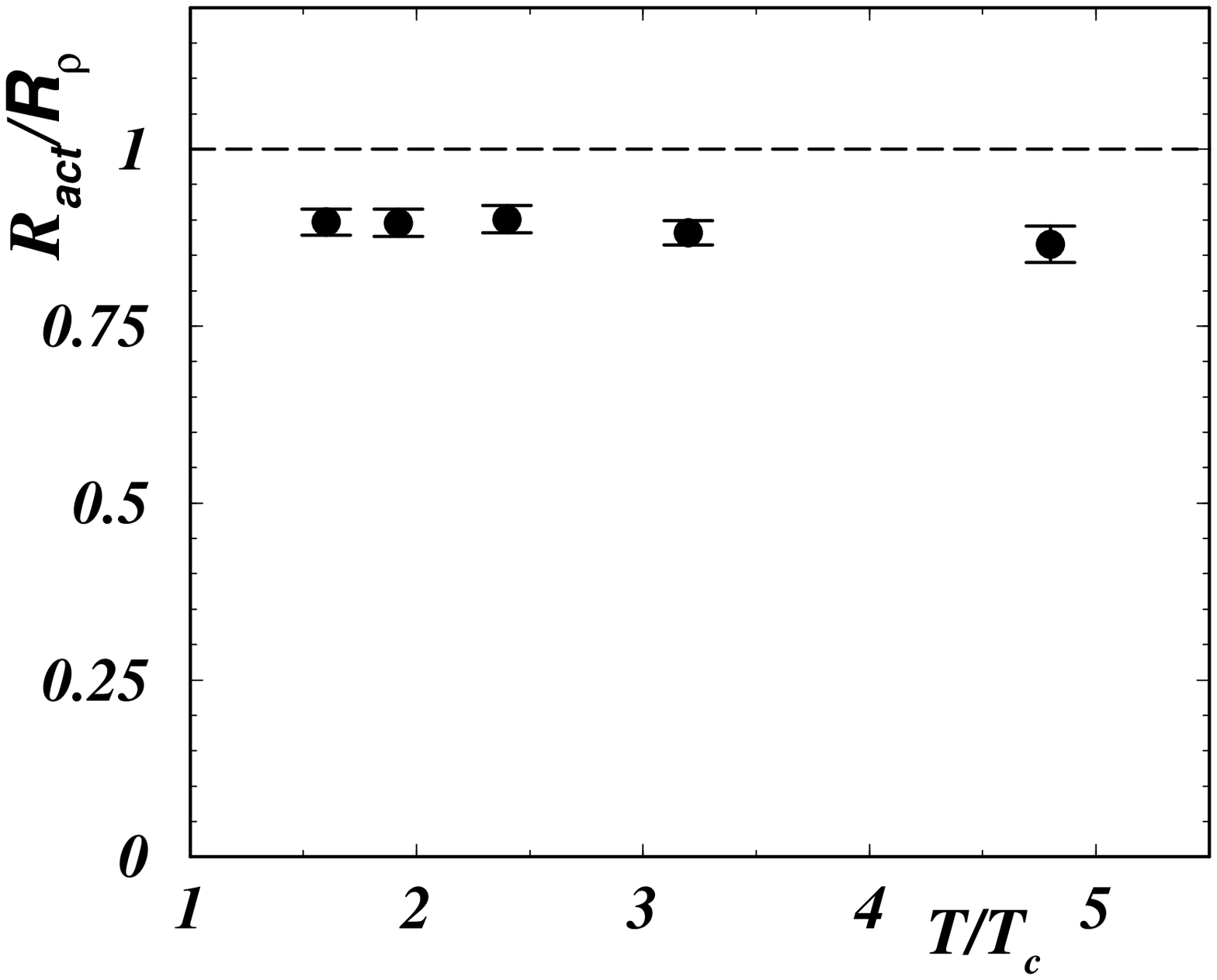,width=5.8cm,height=5.0cm} &
  \epsfig{file=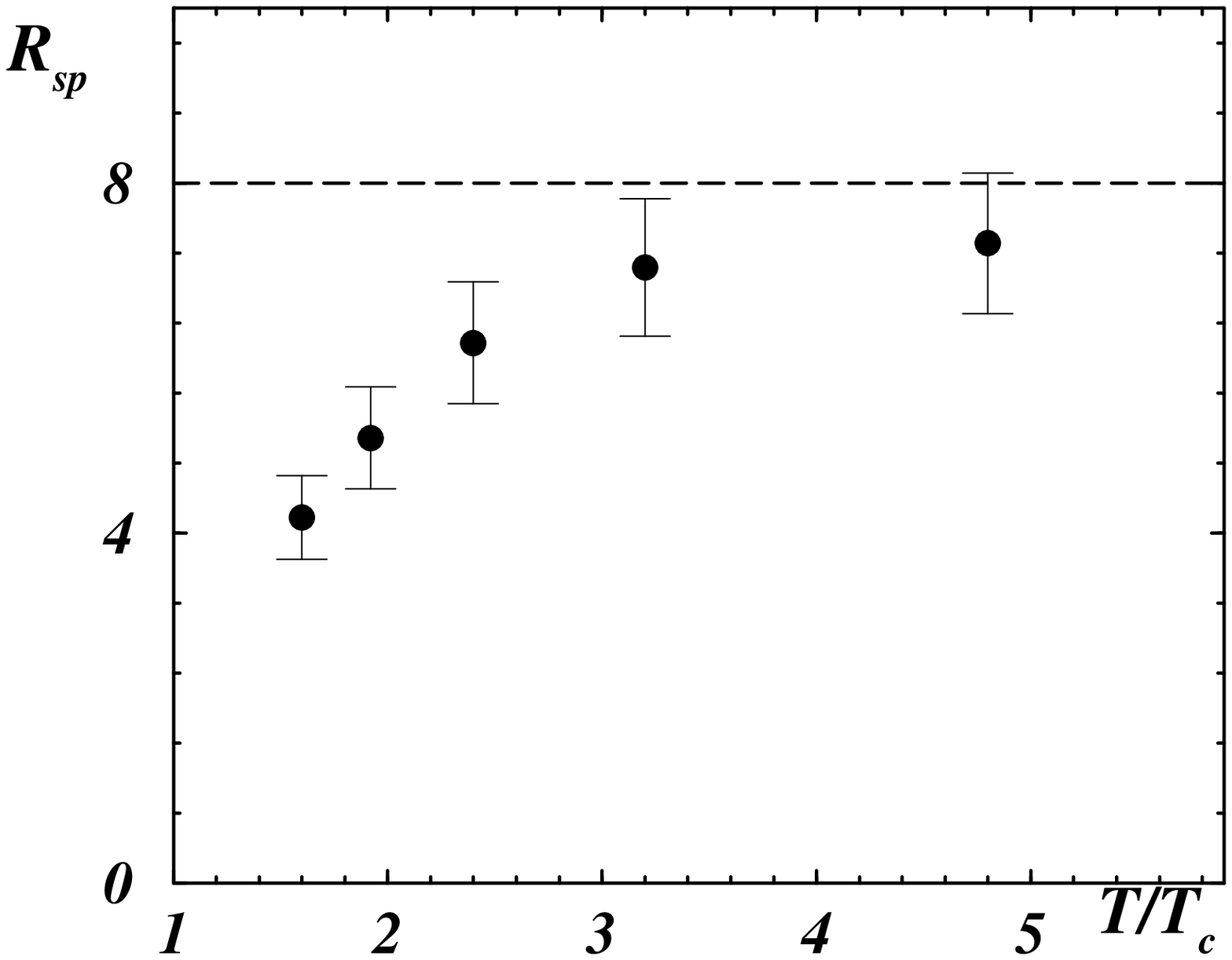,width=5.8cm,height=5.0cm}\\
(a) & (b)\\
\end{tabular}
\end{center}
\caption{(a) Check of self--consistency: the ratio of
the quantities $R$, eq.\eq{R} obtained from
the lattice monopole action and the monopole density;
(b) Check of the dilute Coulomb gas picture: quantity $R_{sp}$, eq.\eq{Rsp},
obtained from the lattice monopole density.}
\label{fig:self}
\end{figure}

In this Section we present our results for the quantity $R$, eq.\eq{R},
which we have obtained both from the large--$b$ behaviour of the
density $\langle k^2(b) \rangle$ and from the monopole action (in these
cases we call the quantity $R$ as $R_\rho$ and $R_{act}$,
respectively). From a numerical point of view the
quantities $R_\rho$ and $R_{act}$ are independent. Thus a natural
condition of a self--consistency of our approach is $R_\rho =
R_{act}$. We check the self--consistency in
Figure~\ref{fig:self}(a) plotting the ratio of these quantities.
It is clearly seen that the ratio is independent on the temperature
and very close to unity, as expected.
The details of our calculations and further discussions will be presented
in Ref.~\cite{ToBePublished}.

Another interesting quantity is the ratio
\beqn
R_{sp}(T) = \frac{\sigma_{sp}(T)}{\lambda_D(T)\, \rho(T)} \equiv
R(T) \cdot \frac{\sigma_{sp}(T)}{\sigma_{T=0}}\,,
\label{Rsp}
\eeqn
where $\sigma_{sp}$ is the spatial string tension. In the dilute Coulomb gas
of mo\-no\-po\-les we have the following relations~\cite{Polyakov}:
\beqn
m_D \equiv \lambda^{-1}_D = g_M \, \sqrt{\rho}\,,\quad
\sigma= \frac{8}{g_M}
\sqrt{\rho}\,,
\eeqn
which imply $R_{sp} = 8$. Using results of Ref.~\cite{SigmaSP}
for the spatial string tension in hot gluodynamics, $\sigma_{sp}(T)$,
we get the quantity $R_{sp}$ as a function of the temperature, $T$,
Figure~\ref{fig:self}(b). If the Coulomb picture works then
$R_{sp}$ should be close to $8$.
{}From Figure~\ref{fig:self}(b) we conclude that this is
indeed the case at sufficiently large temperatures.

\section{Conclusion}

In order to describe the lattice monopole dynamics we have
proposed to consider the lattice as a measuring device for the
continuum monopoles. We have suggested that at high temperatures
the static lattice mo\-no\-po\-les can be described by a continuum
Coulomb gas model. As a result we are able to draw the following
conclusions:

\begin{itemize}

\item The {\it continuum} Coulomb gas model can describe
the results of the Monte Carlo simulations for the density and
action of the {\it lattice} mo\-no\-po\-les. The dependence of
these quantities on the physical sizes of the lattice
mo\-no\-po\-les is in agreement with the analytical predictions.

\item The parameters of the continuum Coulomb gas model obtained
from investigation of the monopole density and action are
consistent with each other and -- at sufficiently high
temperatures -- with known results for the spatial string tension.

\item The lattice monopole action is dominated by the mass and
Coulomb terms for, respectively, small and large {\it lattice}
monopoles.

\item At sufficiently high temperatures the spatial string tension
is dominated by contributions from the continuum static
monopoles.

\end{itemize}

\section*{Acknowledgments}

Authors thank E.--M.~Ilgenfritz, M.I.~Polikarpov, H.~Reinhardt and
V.I.~Zakharov for interesting discussions. M.~N.~Ch. is supported
by the JSPS Fellowship P01023. M.~N.~Ch. thanks the organizers of
the Workshop for the kind hospitality extended to him and for the
miraculous atmosphere in Star\'a Lesn\'a.

\end{document}